\pdfoutput=1

\documentclass[fleqn,usenatbib]{mnras}

\usepackage{newtxtext,newtxmath}
\usepackage{multirow}
\usepackage{ifpdf}

\usepackage[T1]{fontenc}

\DeclareRobustCommand{\VAN}[3]{#2}
\let\VANthebibliography\thebibliography
\def\thebibliography{\DeclareRobustCommand{\VAN}[3]{##3}\VANthebibliography}

\usepackage{graphicx}	
\usepackage{amsmath}	

\usepackage{tikz}
   \usetikzlibrary{positioning,spy}
\usepackage{graphicx}
\usepackage{subfig}
\usepackage{tikz,xcolor,hyperref}

\definecolor{lime}{HTML}{A6CE39}
\DeclareRobustCommand{\orcidicon}{%
	\begin{tikzpicture}
	\draw[lime, fill=lime] (0,0) 
	circle [radius=0.16] 
	node[white] {{\fontfamily{qag}\selectfont \tiny ID}};
	\draw[white, fill=white] (-0.0625,0.095) 
	circle [radius=0.007];
	\end{tikzpicture}
	\hspace{-2mm}
}
\foreach \x in {A, ..., Z}{%
	\expandafter\xdef\csname orcid\x\endcsname{\noexpand\href{https://orcid.org/\csname orcidauthor\x\endcsname}{\noexpand\orcidicon}}
}

\title[A metal-rich cool core lacking a BCG]{The composition and thermal properties of a cool core lacking a brightest cluster galaxy}

\author[Su et al.]{
Yuanyuan Su,$^{1}$\thanks{E-mail:ysu262@g.uky.edu}\orcidA{}
Francoise Combes,$^{2}$\orcidB{}
Valeria Olivares,$^{1}$
Gianluca Castignani,$^{3,4}$\orcidD{}
Pablo Torne$^{5}$
\newauthor
and Reinout van Weeren$^{6}$\orcidF{}
\\
$^{1}$Department of Physics and Astronomy, University of Kentucky, 505 Rose Street, Lexington, KY 40506, USA\\
$^{2}$Observatoire de Paris, LERMA, Coll\`ege de France, CNRS, Sorbonne University, PSL University, 75014 Paris, France\\
$^{3}$Dipartimento di Fisica e Astronomia ``Augusto Righi”, Alma Mater Studiorum Universit\`{a} di Bologna, Via Gobetti 93/2, I-40129
Bologna, Italy\\
$^{4}$INAF - Osservatorio  di  Astrofisica  e  Scienza  dello  Spazio  di  Bologna,  via  Gobetti  93/3,  I-40129,  Bologna,  Italy\\
$^{5}$Institut de Radioastronomie Millim\'{e}trique, Avda. Divina Pastora 7, Local 20, 18012 Granada, Spain\\
$^{6}$Leiden Observatory, Leiden University, PO Box 9513, 2300 RA Leiden, the Netherlands
}

\date{Accepted XXX. Received YYY; in original form ZZZ}

\pubyear{2023}

\begin{document}
\label{firstpage}
\pagerange{\pageref{firstpage}--\pageref{lastpage}}
\maketitle

\begin{abstract}
We present a multiwavelength observation of a cool core that does not appear to be associated with any galaxy, in a nearby cluster, Abell~1142. Its X-ray surface brightness peak of $\lesssim2$\,keV is cooler than the ambient intracluster gas of $\gtrsim3$\,keV, and is offset from its brightest cluster galaxy (BCG) by 80\,kpc in projection, representing the largest known cool core -- BCG separation. This BCG-less cool core allows us to measure the metallicity of a cluster center with a much-reduced contribution from the interstellar medium (ISM) of the BCG. XMM-Newton observation reveals a prominent Fe abundance peak of $1.07^{+0.16}_{-0.15}$\,Z$_{\odot}$ and an $\alpha/$Fe abundance ratio close to the solar ratio, fully consistent with those found at the centers of typical cool core clusters. {This finding hints that BCGs play a limited role in enriching the cluster centers. However, the discussion remains open, given that the $\alpha/$Fe abundance ratios of the orphan cool core and the BCG ISM are not significantly different.}
Abell~1142 may have experienced a major merger more than 100\,Myr ago, which has dissociated its cool core from the BCG. This implies that the Fe abundance peak in cool core clusters can be resilient to cluster mergers. Our recent IRAM 30-m observation did not detect any CO emission at its X-ray peak
and we find no evidence for massive runaway cooling in the absence of recent AGN feedback. The lack of a galaxy may contribute to an inefficient conversion of the ionized warm gas to the cold molecular gas.
\end{abstract}




\section{Introduction}

The intracluster medium (ICM), constituting 90\% of the cluster baryons, is the reservoir of nearly all metals that have ever been produced by member galaxies. The ICM is therefore an ideal and unique laboratory to study the enrichment process of the Universe as well as to constrain models of supernova nucleosynthesis.
The bulk of the ICM of various different clusters tends to have Fe metallicities consistent with a constant value of 0.3 solar \citep{Urban2017}, which supports early enrichment: chemical elements were deposited and mixed into the intergalactic medium (IGM) before clusters formed \citep{Urban2017,Werner2013,Simionescu2015}. However, the uncertainty of these measurements is greater than 20\% and the measurements were performed with a spatial resolution of a few hundred kpc, therefore a non-uniform composition cannot be ruled out. Furthermore, the observed Fe mass in the ICM, when integrated to large radii, far exceeds what can be produced by the visible stellar population \citep{Sarkar2022,Ghizzardi2021, Blackwell2021}, which is difficult to reconcile with the prevailing enrichment model. 


Another outstanding puzzle has arisen in the study of cluster centers. 
Fe metallicity peaks sharply towards cluster centers in cool core (CC) clusters, which is generally considered to be due to the metals accumulated in the interstellar medium (ISM) of the brightest cluster galaxy (BCG). 
The composition of the gas at the cluster centers, as inferred from the abundance ratios, is consistent with that of our solar system \citep[e.g.,][]{Mernier2018}.  
The enrichment of the IGM, before the formation of galaxy clusters, is likely to have a supersolar $\alpha$/Fe ratio, due to the relatively short star
formation timescales for core collapse supernova (SNcc).  
The stellar mass loss from the BCG may have enriched cluster centers with additional type Ia supernova (SNIa) yields, after the gravitational collapse of clusters, as it would take a much longer time to form white dwarfs.
The observed abundance pattern at the cluster centers would require these two sources of metals to compensate for each other to produce the solar ratios for nearly all clusters studied so far. This striking coincidence has been coined as the ``ICM solar composition paradox" \citep{Mernier2018}. 
Fe metallicity also increases towards the center of non cool core (NCC) clusters but with a much smaller gradient, which has been attributed to mergers disrupting the cluster central metallicity peak \citep{Leccardi2010}.

Cluster cool cores also feature a sharp X-ray surface brightness peak, where the gas is expected to cool to below $100$\,K and rapidly form stars at rates of 100-1000 M$_{\odot}$\,yr$^{-1}$, exceeding the observed star formation rates by orders of magnitudes \citep{Peterson2003}.    
Mechanical feedback from the active galactic nucleus (AGN) of the BCG is likely preventing this hot ICM from cooling, as revealed by the ubiquitous X-ray cavities at cluster cool cores
(see \citet{Fabian2012} for a review). 
Recent observations of a high redshift ($z=1.7$) cluster, SpARCS1049+56, have added new insight into this picture \citep{Webb_2017, Castignani2020, Hlavacek-Larrondo2020}. 
Stars are forming at its center, at an enormous rate of 
$860\pm140$\,M$_{\odot}$\,yr$^{-1}$, in association with a large reservoir of molecular gas of $1.1\times10^{11}$\,M$_{\odot}$. 
{The cold gas and star formation can in principle be fueled by massive, runaway cooling of the intracluster gas, perhaps precisely due to the absence of AGN feedback, as its X-ray peak is offset from the BCG.} 
A relevant object was discovered in the nearby Universe --
``an orphan cloud" serendipitously detected on the outskirt of Abell~1367 \citep{Ge2021}: an isolated cloud (no optical counterpart) with an effective radius of 30\,kpc, detected in X-rays, H${\alpha}$, and CO, which may have been stripped off from an unidentified member galaxy \citep{J_chym_2022}.

\begin{figure*}
\centering
\includegraphics[width=16.5cm, height=16.4cm]{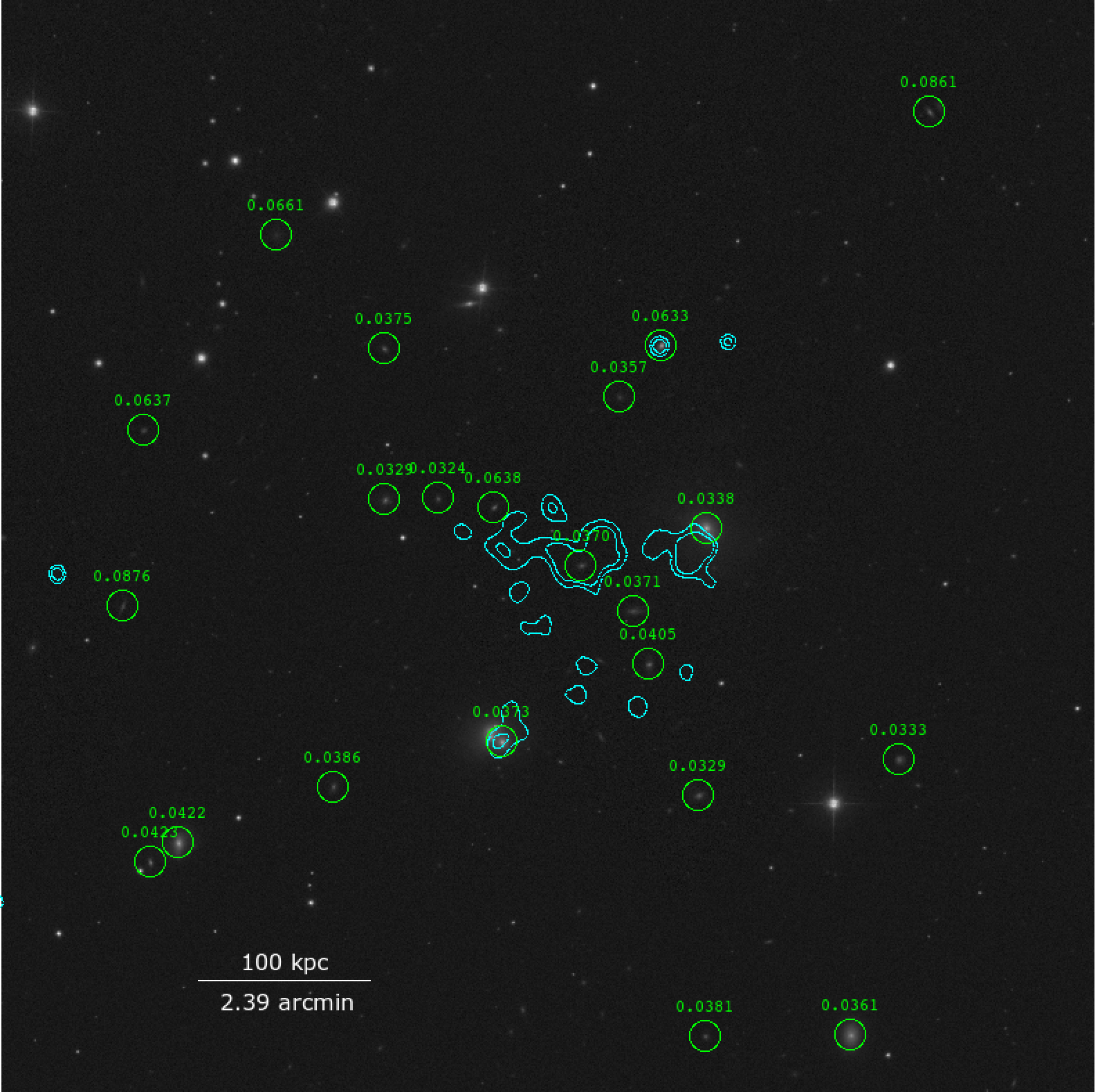}
\caption{{\it SDSS} image of Abell~1142. The cyan contours overlaid trace the {\sl Chandra} X-ray surface brightness as presented in \citet{Su2016}. Spectroscopically confirmed galaxies are circled in green with their spectroscopic redshifts labeled. 
\label{fig:optical}}
\end{figure*}

Abell~1142 is a nearby galaxy cluster that has connections to both SpARCS1049+56 and the orphan cloud. It is undergoing a major merger and the spectroscopically confirmed member galaxies in Abell~1142 display a nearly bimodal redshift distribution peaking at $z=0.0325$ and $0.0375$, respectively (Figure~\ref{fig:optical}; also see \citet{Su2016}). 
 {\sl Chandra} Advanced CCD Imaging Spectrometer (ACIS)-S observation reveals that its X-ray peak is 80\,kpc east of the BCG, IC\,664, a massive elliptical galaxy at $z=0.0338$ \citep{Su2016}. This X-ray emission is unlikely to be a background cluster as this field is covered by SDSS and there is no known optical counterpart at higher redshift \citep{Su2016}.  
Abell~1142 presents the largest known offset between a cluster cool core and the BCG\footnote{Galaxy clusters known to have $>80$\,kpc offset between cool core and BCG are found to be binary or triple clusters, for which the cool core is occupied by the brightest galaxy in the subcluster \citep{De_Propris2021}.}. It provides a valuable opportunity, in the nearby Universe,  
to unambiguously measure the metallicity of a cluster cool core itself, with a minimized impact of the ISM of the BCG, and to test the role of the AGN feedback at cluster centers. 
Previous Chandra observations reveal 
a temperature of 2\,keV and a metal abundance of 1\,Z$_{\odot}$ at the cool core, cooler and more metal rich than the ambient ICM of 3\,keV and 0.2\,Z$_{\odot}$. Notably, its metallicity profile is consistent with that derived for typical cool-core clusters and deviates from that of non cool core clusters (see \citet{Su2016}).

This paper presents follow-up multiwavelength observations of Abell~1142, using XMM-Newton and IRAM 30m, to study the two-dimensional distribution of its gas properties and search for multiphase gas associated with the orphan cool core. 
The paper is structured as follows. 
The data preparation and methods are presented in Section \ref{sec:data_preparation}. The results of the metallicity and multiple phase gas measurements are shown in Section \ref{sec:result}. 
Our findings are discussed in Section \ref{sec:discussion} and summarized in Section \ref{sec:conclusion}.
We use NASA/IPAC Extragalactic 
Database\footnote[3]{\url{http://ned.ipac.caltech.edu}}
to estimate the luminosity distance of 153.9 Mpc 
(1$\arcsec$ = 0.697 kpc) at $z$ = 0.035 for Abell~1142, by adopting a 
cosmology of H$_0$ = 70 km s$^{-1}$ Mpc$^{-1}$, 
$\Omega_{\Lambda}$ = 0.7, 
and $\Omega_\textrm{m}$ = 0.3. 
We assume a solar abundance table of \citet{2009ARA&A..47..481A}
throughout this paper.
All uncertainties are reported at 1$\sigma$ confidence level.

\section{observations and data reduction}
\label{sec:data_preparation}
\begin{table*}
\caption{List of {\sl XMM-Newton} and IRAM-30m observations presented in this paper}
\centering
\begin{tabular}{|c c c c c|} 
 \hline
Instrument & Observation & Date & Effective Exposure& PI \\ [0.5ex] 
 \hline\hline
 XMM-Newton EPIC & 0782330101 & Jun 2016 & 37\,(MOS), 20\,(pn)\,ksec & \multirow{2}{*}{Y.\ Su}\\
  XMM-Newton EPIC & 0782330301& Jun 2016  & 100\,(MOS), 82\,(pn)\,ksec\\
   \hline
  IRAM 30-m & E01-22 & Feb 2023 & 11 hours & Y.\ Su\\ 
 \hline
\end{tabular}
\label{table:1}
\end{table*}

\subsection{XMM-Newton}
Abell~1142 was observed with {\sl XMM-Newton} in 2016 for an effective time of 100\,ksec (Figure~\ref{fig:xmm} and Table~\ref{table:1}). Data reduction was performed with the Science Analysis System (SAS) version xmmsas\_20210317\_1624-19.1.0. ODF files were processed using {\tt emchain} and {\tt epchain}. 
Soft flares from MOS and pn data were filtered using {\tt mos-filter} and {\tt pn-filter}, respectively. Only ${\rm FLAG}=0$ and ${\rm PATTERN}<=12$ event files are included for MOS data and only ${\rm FLAG}=0$ and ${\rm PATTERN}<=4$ event files are included for pn data. Out-of-time pn events were also removed. Point sources were detected by {\tt edetect\_chain}, confirmed by eye, and removed from spectral analysis. 

We extracted spectra from an annulus of 7.4--11.9 arcmin in radius to determine the astrophysical X-ray background (AXB) as well as the non-X-ray background (NXB) simultaneously. 
To better constrain the level of the local bubble and Milky Way foregrounds, we also included a RASS spectrum extracted from an annulus centered on but beyond the field of view of the XMM-Newton pointing. 
The
spectral fit was restricted to the 0.3–10.0 keV energy band for XMM-Newton and 0.1-2.4\,keV for RASS. 
We use {\tt phabs*}({\tt pow}$_{\rm CXB}$+{\tt apec}$_{\rm MW}$)+ {\tt apec}$_{\rm LB}$
to model the AXB. 
A power law {\tt pow}$_{\rm CXB}$ with index $\Gamma=1.41$ represents the cosmic X-ray background (CXB), a thermal emission {\tt apec}$_{\rm LB}$ with a temperature of 0.08\,keV for local bubble emission, and another {\tt apec}$_{\rm MW}$ with a temperature of 0.2\,keV was for the Milky Way foreground. Metal abundance and redshift were fixed at 1 and 0, respectively, for {\tt apec}$_{\rm LB}$ and {\tt apec}$_{\rm MW}$. The aforementioned annulus is chosen as far from the cluster center as possible, while still ensuring enough photons to constrain the background components. Still, it may contain ICM emission. Therefore, we include {\tt apec}$_{\rm ICM}$ to model the ICM component and allow its temperature, abundance, and normalization free to vary. 
The NXB model includes a set of Gaussian lines and a broken powerlaw with E$_{\rm break}=3$\,keV to model a set of fluorescent instrumental lines and a continuum spectrum for each MOS and
pn detector (see the Appendix in \citet{Su2017b} for details). 
The NXB model was
not folded through the Auxiliary Response Files (ARF).
We inspect the ratio of
area-corrected count rates in the 6–12 keV energy band within the field of view (excluding the central 10 arcmin) and in the unexposed corners. Our observations are not contaminated by the soft proton flare.

\subsection{IRAM 30-m}

We performed an observation of Abell~1142 with the IRAM 30-m telescope operated by the Institut de Radio Astronomie Millim\'{e}trique (IRAM) at Pico Veleta, Spain. 
The observation was carried out with the wobbler switching mode (WSW), targeting a few positions\footnote{These positions are chosen based on a tentative detection of extended CO(1-0) emission from a previous IRAM 30m observation using on-the-fly mapping mode (030-22). But we now consider this ``detection" flawed by baseline ripples due to poor weather.} mainly near the X-ray centroid as shown in Figure~\ref{fig:xmm}. The observation (E01-22) was carried out from 21 February 2023 to 22 February 2023 (Table~\ref{table:1}). The EMIR receiver was used to simultaneously observe at the frequencies of the CO(1-0) and CO(2-1) transitions. The FTS spectrometer and the WILMA autocorrelator were connected to both receivers. 1055+018 and IRC+10216 were used as flux calibrators. 

The half power beam width of the IRAM 30m is $\sim22\arcsec$ for CO(1-0) and $\sim11\arcsec$ for CO(2-1). Data reduction was performed using CLASS from the GILDAS\footnote{\url{http://www.iram.fr/IRAMFR/GILDAS}} software package. The corrected antenna temperatures, Ta*, were converted to the brightness temperature of the main beam by Tmb = Ta* F$_{\rm eff}$ / $\eta_{\rm mb}$. The main beam efficiency is $\eta_{\rm mb}$ = 0.78 and 0.61 for 110 GHz and 220 GHz, respectively, and the forward efficiencies F$_{\rm eff}=0.94$ and 0.93 respectively. 
The flux density is converted from the main beam antenna temperature by the ratio $\sim5$\,Jy\,K$^{-1}$ both for CO(1-0) and CO(2-1) transitions. 
However, we did not detect any emission line associated with the CO(1-0) or CO(2-1) transition. 

\section{Results}
\label{sec:result}
\subsection{Metallicity}

\begin{figure*}
\centering
\includegraphics[width=16.5cm, height=9.69cm]{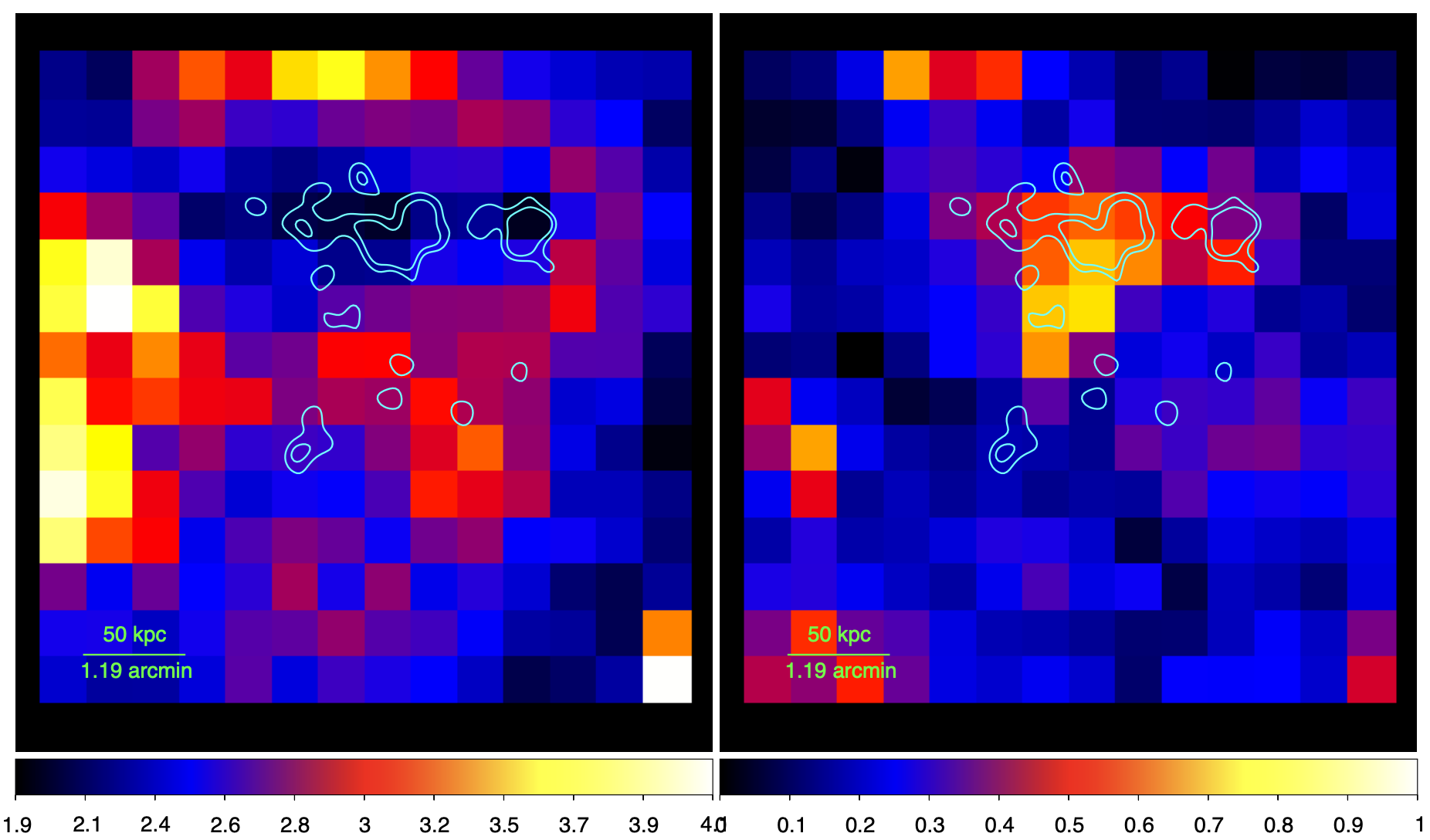}
\caption{
{\it left:} temperature map in units of keV obtained with {\sl XMM-Newton}. {\it right:} metallicity map in units of solar abundance obtained with {\sl XMM-Newton}. Overlaid is the cyan contour tracing the {\sl Chandra} X-ray surface brightness as presented in \citet{Su2016}. 
\label{fig:map}}
\end{figure*}

We produce adaptively binned temperature and metallicity maps
of Abell~1142 using XMM-Newton observations. A square region with a side length of 7.56 arcmin was chosen covering the Abell~1142 centroid, major member galaxies, and their surroundings. The region was divided into a 14x14 grid of pixels. We extract a spectrum from a circular region centered on each pixel. Its radius is chosen adaptively to collect at least 500 net counts. Response files were produced for each spectrum.
We apply the same model used for the annulus region to model the AXB, NXB, and ICM components (see Section~2.1) of each region. 
The parameters of the AXB model and most parameters of the NXB model are fixed to those determined earlier. We allow the normalization of the fluorescent instrumental lines
at 1.49\,keV and 1.75\,keV free to vary for MOS and that of the 1.48\,keV line free to vary for pn. We first fix the ICM metallicity at 0.3\,${\rm Z}_{\odot}$ and obtain a temperature map as shown in Figure~\ref{fig:map}-left. We then allow metallicity free to vary to obtain a metallicity map as shown in Figure~\ref{fig:map}-right. The 2D spatial distribution of its gas properties demonstrates that the orphan cool core is cooler than the surrounding ICM. The metallicity peaks around the orphan cool core, approaching 1\,Z$_{\odot}$, in contrast to the ambient ICM outside the cool core of 0.2\,Z$_{\odot}$. 
{The spectrum extraction regions, with radii ranging from 0.5$^{\prime}$ at the center to 1.3$^{\prime}$ at the outskirt, are chosen through adaptive-binning and overlap with adjacent regions}; the actual temperature and metal abundance contrast are likely to be more pronounced.

To constrain the metallicity of the cool core,
we extract a spectrum from a circular region centered on the X-ray peak with a radius of 25\,kpc (cyan circle in Figure~\ref{fig:xmm}). The spectrum was fit to a two-temperature thermal {\tt phabs*(vapec+vapec)} model (along with multiple background components as described in the previous paragraph), in which all the $\alpha$ elements (O, Ne, S, Si, Al, Ca), mainly the products of SNcc, are linked together, while Fe and Ni, mainly synthesized in SNIa, are linked together. 
One of the two temperatures is free to vary, for which we obtain a best-fit temperature of $1.11^{+0.10}_{-0.08}$\,keV, {while the other temperature is fixed at 3\,keV (the temperature at the outer atmosphere of the ICM as shown in Figure~\ref{fig:map}-left)}.
We obtain a best-fit Fe abundance of $1.07^{+0.16}_{-0.15}$\,Z$_{\odot}$ and an $\alpha/$Fe ratio of $0.95^{+0.28}_{-0.27}$ the Solar ratio, consistent with the Fe abundance peak and abundance ratios measured for typical cool core clusters.
We performed a similar spectral analysis for a circular region of $r<14$\,kpc centered on the BCG using exactly the same model\footnote{Using the point source detection tool {\tt wavdetect}, we identified a point-like source at the very center of the BCG from the existing Chandra ACIS-S observation. We extracted the spectrum from this source with a circular region of $r<1.8\arcsec$ and used an annulus region of $1.8\arcsec<r<2.7\arcsec$ as the local background. The spectrum was fit to an absorbed power law model, for which we obtain a best-fit photon index of $4.1\pm0.5$, which is too soft to be a nuclear source and is likely to be thermal emission. Therefore, we did not include an additional power-law component in the spectral analysis of the BCG.}. We obtain a best-fit Fe abundance of $0.78^{+0.21}_{-0.16}$\,Z$_{\odot}$ and an $\alpha/$Fe ratio of $0.65\pm0.28$ the Solar ratio. {This latter measurement, presumably more representative of the BCG ISM metallicity, is more dominated by the SNIa yields, although consistent within the uncertainties with the composition observed for the orphan cool core}.  

\subsection{Cold molecular gas}

To capture a potential multiphase gas near the X-ray centroid,
we performed an IRAM 30m observation of Abell~1142, using the wobbler-switching (WSW) mode to obtain a ripple-free
baseline and targeting on both CO(1-0) and CO(2-1) transitions. We observed a few positions mainly within the orphan core, and we did not detect any CO emission line with a clean baseline. 
A new issue was raised a few days before the observation. The maximum throw had been limited to 30 arcsec, with no fix on the horizon. If the emission is extended, the signal would be reduced due to the limited maximum throw. 
Based on the WSW observation at its X-ray peak (black annulus in Figure~\ref{fig:xmm}), we set 3$\sigma$ upper limits of 1.1Jy km/s and 1.7Jy km/s on the CO(1-0) and CO(2-1) emission lines, respectively, at a velocity resolution of 300 km/s.  
We calculate the upper limit on the CO luminosity from the standard conversion of
\begin{equation}
L^{\prime}_{\rm CO}=3.25\times10^{7}S_{\rm CO}\Delta v \nu^{-2}_{\rm obs} D^2_L(1+z)^{-3}\,\rm K\,km\,s^{-1}\,pc^2
\end{equation}
where $S_{\rm CO}\Delta v$ is the CO velocity integrated flux for an emission line in Jy\,km\,s$^{-1}$, $\nu_{\rm obs}$ is the observed frequency in GHz, and $D_L$ is the luminosity distance in Mpc. 
The molecular gas mass in H$_2$ is estimated from $M_{\rm H_2}[{\rm M_{\odot}}]=\alpha_{\rm CO}\,L^{\prime}_{\rm CO}\,[\rm K\,km\,s^{-1}\,pc^2]$, where we choose $\alpha_{\rm CO}=4.6$\,M$_\odot$/(K\,km\,s$^{-1}$\,pc$^{2}$) as in our Galaxy \citep{solomonybarrett91}. {This CO--H2 conversion only applies to CO(1-0). A factor of $R_{1J}$ must be added for the CO(J, J-1) transitions and in this case we take $R_{12}=1.2$ for CO(2-1) \citep{Tacconi2018}. 
We obtain $3\sigma$ upper limits on $M_{\rm H_2}$ of $2.8\times10^{8}$\,M$_{\odot}$ and $1.4\times10^{8}$\,M$_{\odot}$ based on CO(1-0) and CO(2-1) measurements, respectively.}

\section{discussions}
\label{sec:discussion}
\subsection{Enrichment process of the ICM}

Extensive XMM-Newton observations of the centers of cool core clusters, i.e., the CHEmical Enrichment RGS cluster Sample (CHEERS) catalog of X-ray bright galaxy groups, clusters and elliptical galaxies, have revealed that the ratios of multiple elements relative to Fe (X/Fe) are consistent with the chemical composition of our solar system, and in excellent agreement with the measurement of the center of the Perseus cluster using the microcalorimeter on board Hitomi \citep{Mernier2018}.

The abundance ratios outside cluster cores have been observed with Suzaku for only a few clusters \citep{Sarkar2022, Simionescu2015}. 
However, unlike cluster centers, agreement on abundance ratios of cluster outskirts has not been reached. 
\citet{Simionescu2015} have presented the results of a Suzaku key project on the nearest galaxy cluster, Virgo. Abundance ratios of Si/Fe, Mg/Fe, and S/Fe out to its virial radius have been reported, which are generally consistent with the Solar value.   
Recently, \citet{Sarkar2022} presented the O/Fe, Si/Fe, Mg/Fe, and S/Fe ratios for 4 low mass clusters out to their virial radii.
The $\alpha$/Fe ratios are found to be consistent with Solar abundances at their centers, but increase outside the cluster centers and reach $2\times$ the Solar value. Interestingly, this latter measurement, with $\alpha$/Fe profiles increasing with radius, is fully consistent with the prediction of the IllustrisTNG cosmological simulation \citep{Sarkar2022}. 

If $\alpha$/Fe is found to be uniform in the ICM, from the centers
to the outskirts, it would imply that the metals throughout the ICM have the same origin, likely from an epoch before the gravitational collapse of the clusters. If the compositions of the cluster outskirts deviate from those of the centers, it favors different enrichment channels. For example, the cluster outskirts may have been enriched at an earlier epoch (SNcc dominated), while the centers may have been enriched later on by the stellar mass loss of the BCG (SNIa dominated), as predicted by IllustrisTNG.

 In this study, we performed abundance measurement for a cluster center, for the first time, without the immediate influence of the innermost atmosphere of the BCG -- minimizing the contribution of BCG to the metallicity of the cluster centers. 
 We cannot individually constrain X/Fe for each element. Instead we linked all the elements that are mainly synthesized in core collapse supernovae. The best-fit $\alpha$/Fe ratio of about the Solar value as well as an abundance peak of Fe of $\sim1$ solar for the BCG-less cool core in Abell~1142 are fully consistent with those found for typical cool cores with a BCG. {This finding hints at a limited role of the BCG in
enriching the cluster centers and producing a solar-like composition}.

A robust comparison of the composition of cluster centers and outskirts would be the key to understanding the enrichment process of the ICM. 
 Both of these aforementioned Suzaku studies have sizable uncertainties for measuring cluster outskirts. 
Observations with future X-ray instruments such as the Line Emission Mapper X-ray Probe (LEM) \citep{Kraft2022} may allow us to convincingly measure the abundance ratios of the cluster outskirts.

\subsection{The impact of mergers on metal-rich cool cores}

The dichotomy of cool core and non cool core clusters has been a lasting puzzle in the study of cluster formation and evolution. 
The Fe abundance peak in NCC clusters is not nearly as pronounced as in CC clusters, as an anticorrelation between entropy and metallicity has been observed at cluster centers \citep{Leccardi2010}.
A cool
core is considered a natural outcome of radiative cooling, and
major mergers may have disrupted cluster cool cores and created
NCC clusters \citep{Su2020}.
 This prevailing theory was challenged by a recent
joint Chandra and SPT analysis of 67 galaxy clusters at $0.3<z<
1.3$, revealing that the fraction of CC clusters has not evolved
over the past 9 Gyr \citep{Ruppin2021}. This would require CC
clusters to be converted to NCC clusters by mergers, at exactly the
same rate as NCC clusters transform to CC clusters by cooling, at
every redshift interval, seriously questioning the role of mergers in
destroying cluster cool cores \citep{Olivares2022}.

In this study, we measure an Fe abundance peak of $\sim1.0$\,Z$_{\odot}$ for the central $r<25$\,kpc of the orphan cool core in Abell~1142. This cool core has clearly been through a recent major merger, as indicated by the bimodal velocity distribution of the member galaxies and the large offset between the X-ray peak and BCG. 
The prominent Fe abundance peak of the orphan cool core implies that the Fe peak of CC clusters may be able to survive some mergers. 


\subsection{The lack of cold gas detection in the orphan cool core}

\begin{figure*}
\centering
\includegraphics[width=10cm]{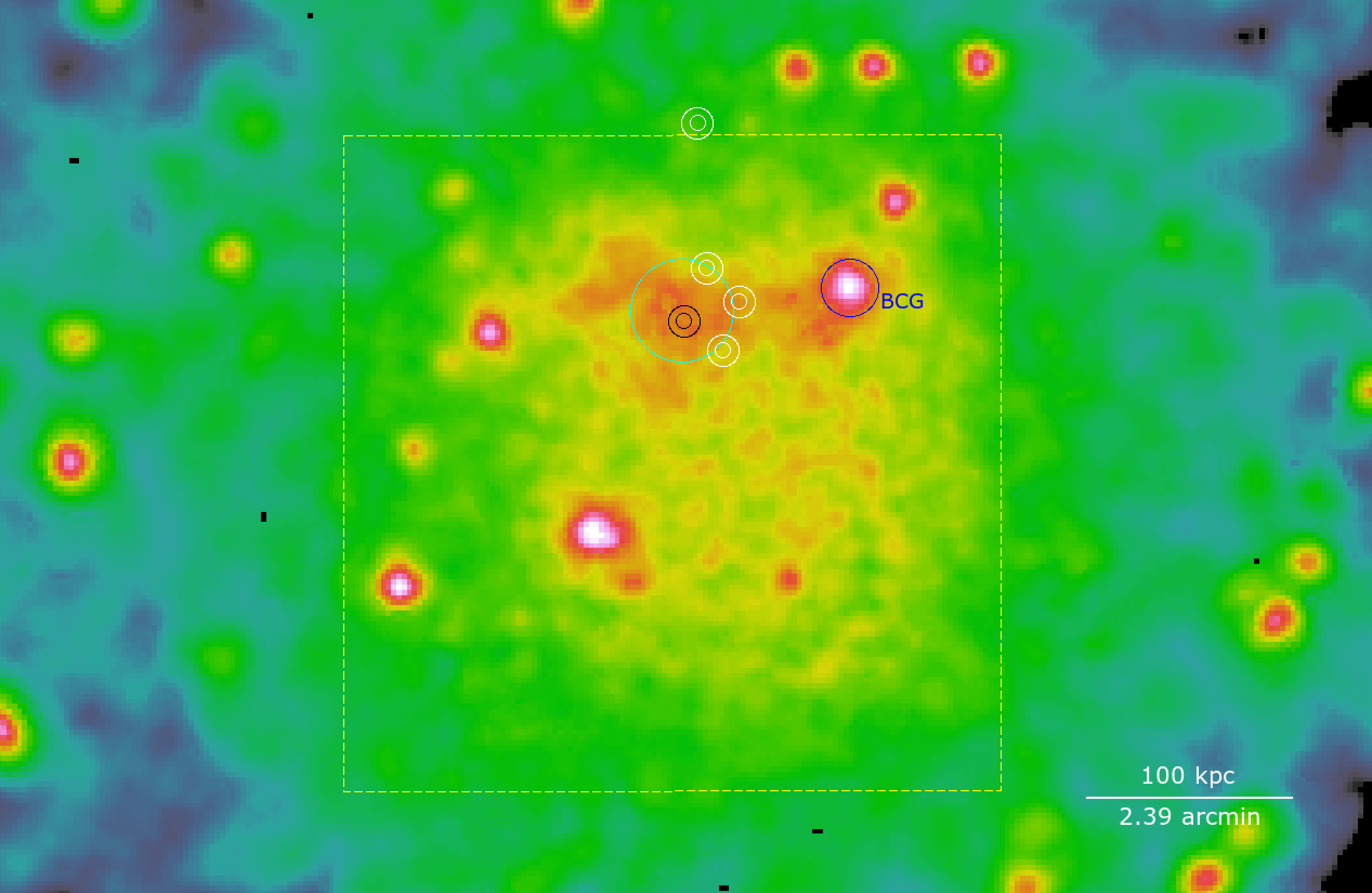}
\caption{Particle background subtracted and vignetting corrected XMM-Newton EPIC image of the Abell~1142 galaxy cluster in the energy band of 0.7--1.3\,keV. The positions of the IRAM 30-m WSW observations are marked in white and black annuli (the inner circle and the outer circle correspond to the beam size of the 230\,GHz and 115\,GHz observations). The regions used to measure the abundance ratio of the cluster center and the BCG are marked in cyan and blue circles. {The yellow dashed square region marks the field of Figure~\ref{fig:map}.}
\label{fig:xmm}}
\end{figure*}

We did not detect cold molecular gas at the X-ray peak of Abell~1142 (black annulus in Figure~\ref{fig:xmm}).
We derive the cooling rate of 1.2\,$M_{\odot}$\,yr$^{-1}$ for the orphan core, using 
\begin{equation}
M_{\rm cf}=\frac{2m_{\mu}m_{\rm p} L_{\rm X}}{5\,\rm kT},
\end{equation}
where $L_{\rm X}=3\times10^{41}$\,erg/s is the bolometric luminosity measured with XMM-Newton and we assume a temperature of 1\,keV. 
The BCG is at a redshift of $0.0338$, which is 800 km/s offset from the median redshift of the member galaxies at $0.0365$. We assume that the BCG travels at 800 km / s relative to the ICM of Abell~1142. Given that the BCG and the orphan cool core are offset by 80\,kpc in projection, we infer that the orphan cool core may have been detached from the BCG for more than 100\,Myr. It would accumulate at least $1.2\times10^{8}$\,M$_{\odot}$ cold gas. The upper limit of $1.4\times10^{8}$\,$M_{\odot}$, obtained from the IRAM 30-m observation, indicates that runaway cooling does not occur in the orphan cool core, even in the absence of AGN feedback. Furthermore, even if cooling takes place as expected, the ionized warm gas may not be converted into cold molecular gas efficiently. 
This conversion ineffectiveness may be precisely due to the lack of a galaxy, as warm and cooling gas usually flows onto the galaxy to form molecular hydrogen, catalyzed by dust. 
A follow-up observation of the ionized hydrogen of Abell~1142 would be critical to resolve the scenarios.

{With low spatial resolution, \citet{Webb_2017} detected $1.1\times10^{11}$\,M$_{\odot}$ of molecular gas,
in the cool core cluster SpARCS1049+56 at $z = 1.7$, which they interpreted to be associated with the X-ray peak. The molecular gas, in this case, could have arisen from massive runaway cooling of the nascent ICM \citep{Hlavacek-Larrondo2020}.
However, the separation between the coolest ICM and BCG is only 25\,kpc in SpARCS1049+56, for which we expect even less time for cooling than Abell~1142. 
In fact, NOEMA observations of SpARCS1049+56  revealed, at high resolution, that 
at least 50\%
of the molecular gas is associated with the BCG companions, and the rest of the cold gas is in a tidal or ram-pressure tail \citep{Castignani2020}. Therefore, most of the cold gas is not associated with the X-ray peak, and cooling cannot be the dominant mechanism for the formation of its cold gas reservoir.} 


It is interesting to compare the orphan cool core in Abell~1142 and the orphan cloud in Abell~1367 \citep{Ge2021}.
Both reside in the nearby Universe and have similar X-ray luminosity ($L_{\rm X,bol}=3\times10^{41}$\,erg/s).
Multiphase gas from X-ray to molecular cold gas has been detected in the orphan cloud, while the orphan cool core is likely to be single phase. 
One possibility is that the cold gas of the orphan cloud, instead of being formed from cooling, is stripped from its host galaxy. The mixing between this stripped cold gas and the ambient hot ICM has led to the
formation of warm ionized gas detected through H$\alpha$ emission. 
Another key difference lies in their hot gas metallicity. 
The orphan cloud has a surprisingly pristine metallicity of 0.14\,Z$_{\odot}$ \citep{Ge2021}, while that of the orphan cool core has a Solar abundance.  
The measured metallicity of the orphan cloud may have been biased low by its complicated temperature structure. 


\section{Conclusions}
\label{sec:conclusion}
In this work, we have presented a multi-wavelength study of the nearby cluster Abell~1142, featuring a cluster cool core that is not associated with any galaxy and offset from its BCG by 80\,kpc. The composition and thermal properties of this orphan cool core can provide unique constraints on the enrichment and cooling processes in the ICM. Our findings obtained with XMM-Newton and IRAM 30-m observations are summarized below. 

\begin{itemize}

\item The central $r<25$\,kpc of the orphan cool core has a prominent Fe abundance peak of $\sim1$ solar abundance. Its $\alpha$/Fe ratio is fully consistent with the solar ratio. Both the metallicity and chemical composition of this orphan cool core (with a reduced impact from the innermost atmosphere of the BCG) are in excellent agreement with those observed for typical cluster cool cores. This result hints that the stellar mass loss of the BCG may have a limited contribution to the enrichment of cluster centers. However, 
the consistent abundance ratios within $1\sigma$ error of the orphan cool core ($0.95\pm{0.28}$ Solar ratio) and the BCG ISM ($0.65\pm0.28$ Solar ratio)
{mean that the issue remains unresolved.}

     \item Abell~1142 has clearly been through a recent major merger that has disassociated the cool core from the BCG and created a bimodal velocity distribution of the member galaxies. The prominent Fe abundance peak, found to be associated with the orphan cool core, indicates that major mergers can not easily erase the Fe peak in cool core clusters.
     
      \item The lack of CO(1-0) and CO(2-1) emission indicates that the orphan cool core is not multiphase. Assuming that it has been detached from the BCG for 100\,Myr, we expect it to accumulate $1\times10^{8}$\,M$_{\odot}$ molecular gas in H$_2$, which is comparable to the upper limit we obtained from the IRAM 30-m observations. An occurrence of runaway cooling in the absence of AGN feedback can be ruled out. 
      The ionized warm gas, {if there is any}, may not be converted into cold molecular gas efficiently because of the lack of a galaxy. 
\end{itemize}

\section*{Acknowledgements}
The authors thank the anonymous reviewer for their helpful comments on the manuscript.
Y.S. and V.O. acknowledge support by NSF grant 2107711,  Chandra X-ray Observatory grants GO1-22126X, GO2-23120X, G01-22104X, NASA grants 80NSSC21K0714 and 80NSSC22K0856. GC acknowledges the support from the grant ASI n.2018-23-HH.0. RJvW acknowledges support from the ERC Starting Grant ClusterWeb 804208. 

\section*{Data Availability}
{The data underlying this article will be shared
on reasonable request to the corresponding author.}



\bibliographystyle{mnras}
\bibliography{mnras_v2} 

\begin{thebibliography}{}
\makeatletter
\relax
\def\mn@urlcharsother{\let\do\@makeother \do\$\do\&\do\#\do\^\do\_\do\%\do\~}
\def\mn@doi{\begingroup\mn@urlcharsother \@ifnextchar [ {\mn@doi@}
  {\mn@doi@[]}}
\def\mn@doi@[#1]#2{\def\@tempa{#1}\ifx\@tempa\@empty \href
  {http://dx.doi.org/#2} {doi:#2}\else \href {http://dx.doi.org/#2} {#1}\fi
  \endgroup}
\def\mn@eprint#1#2{\mn@eprint@#1:#2::\@nil}
\def\mn@eprint@arXiv#1{\href {http://arxiv.org/abs/#1} {{\tt arXiv:#1}}}
\def\mn@eprint@dblp#1{\href {http://dblp.uni-trier.de/rec/bibtex/#1.xml}
  {dblp:#1}}
\def\mn@eprint@#1:#2:#3:#4\@nil{\def\@tempa {#1}\def\@tempb {#2}\def\@tempc
  {#3}\ifx \@tempc \@empty \let \@tempc \@tempb \let \@tempb \@tempa \fi \ifx
  \@tempb \@empty \def\@tempb {arXiv}\fi \@ifundefined
  {mn@eprint@\@tempb}{\@tempb:\@tempc}{\expandafter \expandafter \csname
  mn@eprint@\@tempb\endcsname \expandafter{\@tempc}}}

\bibitem[\protect\citeauthoryear{{Asplund}, {Grevesse}, {Sauval}  \&
  {Scott}}{{Asplund} et~al.}{2009}]{2009ARA&A..47..481A}
{Asplund} M.,  {Grevesse} N.,  {Sauval} A.~J.,   {Scott} P.,  2009, \mn@doi
  [\araa] {10.1146/annurev.astro.46.060407.145222}, \href
  {https://ui.adsabs.harvard.edu/abs/2009ARA&A..47..481A} {47, 481}

\bibitem[\protect\citeauthoryear{{Castignani}, {Combes}  \&
  {Salom{\'e}}}{{Castignani} et~al.}{2020}]{Castignani2020}
{Castignani} G.,  {Combes} F.,   {Salom{\'e}} P.,  2020, \mn@doi [\aap]
  {10.1051/0004-6361/201937155}, \href
  {https://ui.adsabs.harvard.edu/abs/2020A&A...635L..10C} {635, L10}

\bibitem[\protect\citeauthoryear{{De Propris} et~al.,}{{De Propris}
  et~al.}{2021}]{De_Propris2021}
{De Propris} R.,  et~al., 2021, \mn@doi [\mnras] {10.1093/mnras/staa3286},
  \href {https://ui.adsabs.harvard.edu/abs/2021MNRAS.500..310D} {500, 310}

\bibitem[\protect\citeauthoryear{{E Blackwell}, {Bregman}  \& {Snowden}}{{E
  Blackwell} et~al.}{2021}]{Blackwell2021}
{E Blackwell} A.,  {Bregman} J.~N.,   {Snowden} S.~L.,  2021, arXiv e-prints,
  \href {https://ui.adsabs.harvard.edu/abs/2021arXiv210504638E} {p.
  arXiv:2105.04638}

\bibitem[\protect\citeauthoryear{{Fabian}}{{Fabian}}{2012}]{Fabian2012}
{Fabian} A.~C.,  2012, \mn@doi [\araa] {10.1146/annurev-astro-081811-125521},
  \href {https://ui.adsabs.harvard.edu/abs/2012ARA&A..50..455F} {50, 455}

\bibitem[\protect\citeauthoryear{{Ge} et~al.,}{{Ge} et~al.}{2021}]{Ge2021}
{Ge} C.,  et~al., 2021, \mn@doi [\mnras] {10.1093/mnras/stab1569}, \href
  {https://ui.adsabs.harvard.edu/abs/2021MNRAS.505.4702G} {505, 4702}

\bibitem[\protect\citeauthoryear{{Ghizzardi} et~al.,}{{Ghizzardi}
  et~al.}{2021}]{Ghizzardi2021}
{Ghizzardi} S.,  et~al., 2021, \mn@doi [\aap] {10.1051/0004-6361/202038501},
  \href {https://ui.adsabs.harvard.edu/abs/2021A&A...646A..92G} {646, A92}

\bibitem[\protect\citeauthoryear{{Hlavacek-Larrondo}
  et~al.,}{{Hlavacek-Larrondo} et~al.}{2020}]{Hlavacek-Larrondo2020}
{Hlavacek-Larrondo} J.,  et~al., 2020, \mn@doi [\apjl]
  {10.3847/2041-8213/ab9ca5}, \href
  {https://ui.adsabs.harvard.edu/abs/2020ApJ...898L..50H} {898, L50}

\bibitem[\protect\citeauthoryear{J{\'{a}}chym et~al.,}{J{\'{a}}chym
  et~al.}{2022}]{J_chym_2022}
J{\'{a}}chym P.,  et~al., 2022, \mn@doi [Astronomy {\&}
  Astrophysics] {10.1051/0004-6361/202142791}, 658, L5

\bibitem[\protect\citeauthoryear{{Kraft} et~al.,}{{Kraft}
  et~al.}{2022}]{Kraft2022}
{Kraft} R.,  et~al., 2022, \mn@doi [arXiv e-prints]
  {10.48550/arXiv.2211.09827}, \href
  {https://ui.adsabs.harvard.edu/abs/2022arXiv221109827K} {p. arXiv:2211.09827}

\bibitem[\protect\citeauthoryear{{Leccardi}, {Rossetti}  \&
  {Molendi}}{{Leccardi} et~al.}{2010}]{Leccardi2010}
{Leccardi} A.,  {Rossetti} M.,   {Molendi} S.,  2010, \mn@doi [\aap]
  {10.1051/0004-6361/200913094}, \href
  {https://ui.adsabs.harvard.edu/abs/2010A&A...510A..82L} {510, A82}

\bibitem[\protect\citeauthoryear{{Mernier} et~al.,}{{Mernier}
  et~al.}{2018}]{Mernier2018}
{Mernier} F.,  et~al., 2018, \mn@doi [\mnras] {10.1093/mnrasl/sly134}, \href
  {https://ui.adsabs.harvard.edu/abs/2018MNRAS.480L..95M} {480, L95}

\bibitem[\protect\citeauthoryear{{Olivares}, {Su}, {Nulsen}, {Kraft},
  {Somboonpanyakul}, {Andrade-Santos}, {Jones}  \& {Forman}}{{Olivares}
  et~al.}{2022}]{Olivares2022}
{Olivares} V.,  {Su} Y.,  {Nulsen} P.,  {Kraft} R.,  {Somboonpanyakul} T.,
  {Andrade-Santos} F.,  {Jones} C.,   {Forman} W.,  2022, \mn@doi [\mnras]
  {10.1093/mnrasl/slac096}, \href
  {https://ui.adsabs.harvard.edu/abs/2022MNRAS.516L.101O} {516, L101}

\bibitem[\protect\citeauthoryear{{Peterson}, {Kahn}, {Paerels}, {Kaastra},
  {Tamura}, {Bleeker}, {Ferrigno}  \& {Jernigan}}{{Peterson}
  et~al.}{2003}]{Peterson2003}
{Peterson} J.~R.,  {Kahn} S.~M.,  {Paerels} F.~B.~S.,  {Kaastra} J.~S.,
  {Tamura} T.,  {Bleeker} J.~A.~M.,  {Ferrigno} C.,   {Jernigan} J.~G.,  2003,
  \mn@doi [\apj] {10.1086/374830}, \href
  {https://ui.adsabs.harvard.edu/abs/2003ApJ...590..207P} {590, 207}

\bibitem[\protect\citeauthoryear{{Ruppin}, {McDonald}, {Bleem}, {Allen},
  {Benson}, {Calzadilla}, {Khullar}  \& {Floyd}}{{Ruppin}
  et~al.}{2021}]{Ruppin2021}
{Ruppin} F.,  {McDonald} M.,  {Bleem} L.~E.,  {Allen} S.~W.,  {Benson} B.~A.,
  {Calzadilla} M.,  {Khullar} G.,   {Floyd} B.,  2021, \mn@doi [\apj]
  {10.3847/1538-4357/ac0bba}, \href
  {https://ui.adsabs.harvard.edu/abs/2021ApJ...918...43R} {918, 43}

\bibitem[\protect\citeauthoryear{{Sarkar}, {Su}, {Truong}, {Randall},
  {Mernier}, {Gastaldello}, {Biffi}  \& {Kraft}}{{Sarkar}
  et~al.}{2022}]{Sarkar2022}
{Sarkar} A.,  {Su} Y.,  {Truong} N.,  {Randall} S.,  {Mernier} F.,
  {Gastaldello} F.,  {Biffi} V.,   {Kraft} R.,  2022, \mn@doi [\mnras]
  {10.1093/mnras/stac2416}, \href
  {https://ui.adsabs.harvard.edu/abs/2022MNRAS.516.3068S} {516, 3068}

\bibitem[\protect\citeauthoryear{{Simionescu}, {Werner}, {Urban}, {Allen},
  {Ichinohe}  \& {Zhuravleva}}{{Simionescu} et~al.}{2015}]{Simionescu2015}
{Simionescu} A.,  {Werner} N.,  {Urban} O.,  {Allen} S.~W.,  {Ichinohe} Y.,
  {Zhuravleva} I.,  2015, \mn@doi [\apjl] {10.1088/2041-8205/811/2/L25}, \href
  {https://ui.adsabs.harvard.edu/abs/2015ApJ...811L..25S} {811, L25}

\bibitem[\protect\citeauthoryear{{Solomon} \& {Barrett}}{{Solomon} \&
  {Barrett}}{1991}]{solomonybarrett91}
{Solomon} P.~M.,  {Barrett} J.~W.,  1991, in {Combes} F.,  {Casoli} F.,  eds,
  IAU Symposium Vol. 146, Dynamics of Galaxies and Their Molecular Cloud
  Distributions. p.~235

\bibitem[\protect\citeauthoryear{{Su}, {Buote}, {Gastaldello}  \& {van
  Weeren}}{{Su} et~al.}{2016}]{Su2016}
{Su} Y.,  {Buote} D.~A.,  {Gastaldello} F.,   {van Weeren} R.,  2016, \mn@doi
  [\apj] {10.3847/0004-637X/821/1/40}, \href
  {https://ui.adsabs.harvard.edu/abs/2016ApJ...821...40S} {821, 40}

\bibitem[\protect\citeauthoryear{{Su}, {Nulsen}, {Kraft}, {Forman}, {Jones},
  {Irwin}, {Randall}  \& {Churazov}}{{Su} et~al.}{2017}]{Su2017b}
{Su} Y.,  {Nulsen} P. E.~J.,  {Kraft} R.~P.,  {Forman} W.~R.,  {Jones} C.,
  {Irwin} J.~A.,  {Randall} S.~W.,   {Churazov} E.,  2017, \mn@doi [\apj]
  {10.3847/1538-4357/aa8954}, \href
  {https://ui.adsabs.harvard.edu/abs/2017ApJ...847...94S} {847, 94}

\bibitem[\protect\citeauthoryear{{Su} et~al.,}{{Su} et~al.}{2020}]{Su2020}
{Su} Y.,  et~al., 2020, \mn@doi [\mnras] {10.1093/mnras/staa2690}, \href
  {https://ui.adsabs.harvard.edu/abs/2020MNRAS.498.5620S} {498, 5620}

\bibitem[\protect\citeauthoryear{{Tacconi} et~al.,}{{Tacconi}
  et~al.}{2018}]{Tacconi2018}
{Tacconi} L.~J.,  et~al., 2018, \mn@doi [\apj] {10.3847/1538-4357/aaa4b4},
  \href {https://ui.adsabs.harvard.edu/abs/2018ApJ...853..179T} {853, 179}

\bibitem[\protect\citeauthoryear{{Urban}, {Werner}, {Allen}, {Simionescu}  \&
  {Mantz}}{{Urban} et~al.}{2017}]{Urban2017}
{Urban} O.,  {Werner} N.,  {Allen} S.~W.,  {Simionescu} A.,   {Mantz} A.,
  2017, \mn@doi [\mnras] {10.1093/mnras/stx1542}, \href
  {https://ui.adsabs.harvard.edu/abs/2017MNRAS.470.4583U} {470, 4583}

\bibitem[\protect\citeauthoryear{Webb et~al.,}{Webb et~al.}{2017}]{Webb_2017}
Webb T. M.~A.,  et~al., 2017, \mn@doi [The Astrophysical Journal]
  {10.3847/2041-8213/aa7749}, 844, L17

\bibitem[\protect\citeauthoryear{{Werner}, {Urban}, {Simionescu}  \&
  {Allen}}{{Werner} et~al.}{2013}]{Werner2013}
{Werner} N.,  {Urban} O.,  {Simionescu} A.,   {Allen} S.~W.,  2013, \mn@doi
  [\nat] {10.1038/nature12646}, \href
  {https://ui.adsabs.harvard.edu/abs/2013Natur.502..656W} {502, 656}

\makeatother
\end{thebibliography}





\bsp	
\label{lastpage}
\end{document}